\documentclass[aps,epsf]{iopart}
\pdfoutput=1
\usepackage{graphics}
\usepackage{graphicx}
\usepackage{epsfig}

\def\lesssim{\mathrel{\hbox{\rlap{\hbox{\lower4pt\hbox{$\sim$}}}\hbox{$<$}}}}
\def\beq{\begin{equation}}
\def\eeq{\end{equation}}
\def\bea{\begin{eqnarray}}
\def\eea{\end{eqnarray}}

\begin{document}
\input epsf.tex
 
\title[A Constrained MHMC search for EMRIs in the MLDC Round 1B]{A Constrained Metropolis-Hastings Search for EMRIs in the Mock LISA Data Challenge 1B}
\author{Jonathan R. Gair$^1$,  Edward Porter$^2$, Stanislav Babak$^2$ and Leor Barack$^3$} 
\address{(1) Institute of Astronomy, University of Cambridge, Cambridge, CB3 0HA, UK\\(2) Albert Einstein Institute,  Am Muehlenberg 1, D-14476 Golm, Germany\\(3) School of Mathematics, University of Southampton, Southampton, SO17 1BJ, UK}
\ead{jgair@ast.cam.ac.uk}
\vspace{1cm}
\begin{abstract}
\noindent We describe a search for the extreme-mass-ratio inspiral sources in the Round 1B Mock LISA Data Challenge data sets. The search algorithm is a Monte-Carlo search based on the Metropolis-Hastings algorithm, but also incorporates simulated, thermostated and time annealing, plus a harmonic identification stage designed to reduce the chance of the chain locking onto secondary maxima. In this paper, we focus on describing the algorithm that we have been developing. We give the results of the search of the Round 1B data, although parameter recovery has improved since that deadline. Finally, we describe several modifications to the search pipeline that we are currently investigating for incorporation in future searches.
\end{abstract}

\maketitle

\section{Introduction}
The inspirals of stellar mass compact objects into supermassive black holes in the centres of galaxies --- extreme-mass-ratio inspirals (EMRIs) --- are one of the most exciting potential sources of gravitational waves for the planned Laser Interferometer Space Antenna (LISA). The detection of such sources in the LISA data stream and parameter estimation for them is a very challenging technical problem, however, as the instantaneous amplitude of a typical signal is an order of magnitude below the noise fluctuations in the detector. Moreover, the long duration of the signals (LISA will detect up to $10^5$ waveform cycles in an observation) and the large parameter space of possible sources (an EMRI signal depends on fourteen parameters) makes fully-coherent matched filtering computationally impossible~\cite{gairetal}. 

Several possible algorithms have been considered for EMRI detection. In order to compute initial event rate estimates, a semi-coherent algorithm was proposed, in which the data stream would be divided up into short (two or three week) segments, that would be searched coherently via matched filtering. The signal-to-noise ratio (SNR) would then be built up in a second stage via incoherent summation of power along trajectories through the coherent segments~\cite{gairetal}. This algorithm was designed to make full use of available computing power and its effectiveness has not yet been demonstrated practically. Time-frequency techniques have also been explored~(\cite{wengair05}--\cite{proc13}) and have been shown to be able to both detect and recover parameters~\cite{proc13} when used to search for single, high-SNR EMRIs in instrumental noise. Their effectiveness is likely to be significantly reduced when confronted with more realistic situations, in which there are multiple sources overlapping in time and frequency. However, these algorithms may provide a useful first step in a hierarchical search.

Markov Chain Monte Carlo (MCMC) methods may provide a way to perform matched filtering more efficiently, without requiring a large template bank of possible signals. We follow the convention in the literature and call our search a Metropolis-Hastings Monte Carlo (MHMC) search rather than an MCMC search since it is not actually Markovian  (see Section~\ref{MHMCalg}). Both MCMC and MHMC methods have been explored by various groups in the context of LISA data analysis and have been shown to be very effective when searching for toy models~\cite{andrieu,umstat}, for white-dwarf binaries~\cite{cc05,cc2} and for single or multiple supermassive black hole binaries~(\cite{en3}--\cite{en1}). The use of an MCMC technique for EMRI searches was explored by Stroeer et al.~\cite{stroeer06}, based on a highly simplified model of the EMRI waveform. In the context of the Mock LISA Data Challenges (MLDCs)~\cite{mldc1}, monte carlo methods have been used by ourselves and one other group~\cite{neilemri} to search for the EMRI sources in the Round 2~\cite{mldc2} and Round 1B~\cite{mldc3} data sets. In this paper, we describe the algorithm that we have developed for the EMRI searches, and the performance of the algorithm on the Round 1B challenge data sets. Our search code was adapted from the search code developed by Cornish and Porter~\cite{en3, en2, en4, en1} for supermassive black hole binary searches, but we have incorporated a significant number of refinements that are specific to the EMRI problem.

The paper is organised as follows. In Section~\ref{method} we describe the search algorithm, including a description of the waveform model, the Metropolis-Hastings search engine and various refinements we have tried for the EMRI problem. In Section~\ref{results} we present the results that we had at the time of the MLDC Round 1B deadline (December 2007) and compare these to the true source parameters. Finally, in Section~\ref{fut} we discuss planned future refinements of the search algorithm.

\section{Search algorithm}\label{method}
\subsection{Waveform model}
The EMRI sources in the MLDC releases to date were constructed using the {\it analytic kludge} (AK) model of Barack and Cutler~\cite{BC04}. In this model, the gravitational waveforms describe emission from a Keplerian orbit, but with perihelion and orbital-plane precessions imposed by precessing the observer about the source with rates taken from post-Newtonian expressions. The orbital parameters are also evolved over time using post-Newtonian prescriptions to account for radiation reaction. The EMRI waveforms have emission at multiple frequencies, corresponding to harmonics of the fundamental frequencies of the orbit --- harmonics of the orbital frequency, $\nu$, arise from the eccentricity of the orbit, $e$; harmonics of the perihelion precession rate, $\dot{\gamma}/2\pi$, arise from this precession; and harmonics of the orbital-plane precession rate, $\dot{\alpha}/2\pi$, are present due to the inclination, $\lambda$, of the orbital plane relative to the equatorial plane of the black hole. The frequency of a given waveform harmonic is given by three integers, $(n, l, k)$, as $f = n\nu + l \dot{\gamma}/2\pi + k \dot{\alpha}/2\pi$. The AK waveforms are purely quadrupole in nature, and so $|l|, |k| \leq 2$, but $n$ is unrestricted. However, the eccentricity at plunge of the MLDC sources is limited to $[0.15,0.25]$, and so harmonics with $n \geq 6$ are generally weak.

In our search code, to speed up waveform evaluation, we use a truncated version of the AK model. We include only the $n \leq 5$ and $l=2$ harmonics in the waveform model (harmonics with $l \neq 2$ are significantly suppressed relative to the $l=2$ harmonics). In addition, we expand the Bessel functions that appear in the model~\cite{BC04} in powers of eccentricity, and keep only the three leading terms in the expansion of $J_n(ne)$ for each $n$. We include the full LISA TDI response function to account for detector motions, and use parameters evaluated at plunge to characterize the waveforms (this is in contrast to the MLDC convention, which is to specify parameters at the start of the observation). The resulting waveforms are faithful approximations to the full AK waveforms. The overlap between an AK and truncated waveform, evaluated at the same waveform parameters, is at worst $90\%$, and is typically $95\%$. The overlap tends to be higher for sources with higher mass central black holes, for which the emission is mostly at lower frequencies (for $M \sim 10^7M_{\odot}$ the overlap always exceeds $96\%$). The template parameter error, i.e., the difference between the parameters of the best-fit truncated waveform and the parameters of the true AK waveform, is also relatively small. In Table~\ref{parambias} we list the parameter errors for one of the MLDC Round 2 training sources. These results were obtained by starting an MCMC chain at the true parameter values and allowing it to evolve to evaluate the posterior. The difference between the mean of the recovered posterior and the true parameter values provided an estimate of the bias in our truncated model.  We note that since the data stream we were searching included noise we expected and saw a noise-induced bias in the parameter estimation.  However, we could distinguish this noise bias from the model errors.  The parameter offset values in Table~\ref{parambias} are typical for MLDC type sources. The parameters are the same as those used for the MLDC --- compact object mass ($m$), central black hole mass ($M$), initial orbital frequency ($\nu_0$), luminosity distance ($D_L$), initial eccentricity ($e_0$), central black hole spin ($\chi$), ecliptic latitude and longitude ($\beta$, $\phi_S$), orientation of central black hole spin ($\theta_K$, $\phi_K$), orbital inclination ($\lambda$) and three initial orbital phases ($\Phi_0$, $\gamma_0$, $\alpha_0$). For most parameters, the error is at most $2\sigma$, where $\sigma$ is the noise-induced uncertainty in the parameter, as estimated from the width of the posterior. The error is somewhat larger for the initial phase angles, as the effect of the truncation accumulates over the observation, but these parameters are uninteresting astrophysically. Overall, the truncated model provides an estimate of the parameters that is sufficiently close to the true parameters to ensure that a follow-up refinement with full AK waveforms would be quick.

\begin{table}
\begin{tabular}{|c|c|c|c|c|c|c|c|}
\hline
Parameter &$\ln(m/M_{\odot})$&$\ln(M/M_{\odot})$&$\ln(\nu_0/{\rm Hz})$&$D_L/{\rm Gpc}$&$e_0$&$\chi$&$\cos(\lambda)$\\\hline
Error &0.006\%&0.0008\%&$0.00004\%$&7\%&0.07\%&0.006\%&0.06\% \\\hline\hline
Parameter &$\cos(\beta)$&$\phi_S$&$\cos(\theta_K)$&$\phi_K$&$\Phi_0$&$\gamma_0$&$\alpha_0$\\\hline
Error &0.2\%&0.06\%&0.06\%&0.3\%&3\%&3\%&0.5\% \\\hline\hline
\end{tabular}
\caption{Percentage difference between the parameters of the best-fit truncated waveform and the parameters of the true AK waveform. The true AK waveform is the 1.3.2 training waveform from MLDC Round 2, for which the parameters are $\ln(m/M_{\odot})=2.3338$, $\ln(M/M_{\odot})=15.421$, $\ln(\nu_0/{\rm Hz})=-7.9399$, $D_L/{\rm Gpc}=0.80533$, $e_0=0.18765$, $\chi=0.68497$, $\cos(\lambda)=-0.43960$, $\cos(\beta)=0.96737$, $\phi_S=3.6238$, $\cos(\theta_K)=-0.55434$, $\phi_K=4.3216$, $\Phi_0=3.3913$, $\gamma_0=6.1502$ and $\alpha_0=3.2400$.}
\label{parambias}
\end{table}

\subsection{Metropolis-Hastings Monte Carlo algorithm}
\label{MHMCalg}
Our search engine is based in the Metropolis-Hastings algorithm, which works as follows: Given a data set $s(t)$ and a set of templates $h(t;\vec{x})$, we randomly choose a starting point, $\vec{x}$, in the parameter space.  We then propose a jump to another point, $\vec{y}$, in the space by drawing from a certain proposal distribution, $q(\vec{y}|\vec{x})$ (see below), and evaluate the Metropolis-Hastings ratio
\begin{equation}
H = \frac{\pi(\vec{y})p(s|\vec{y})q(\vec{x}|\vec{y})}{\pi(\vec{x})p(s|\vec{x})q(\vec{y}|\vec{x})}.
\end{equation}
Here $\pi(\vec{x})$ are the priors of the parameters, which, in our analysis, were taken to be uniform distributions within the ranges allowed by the MLDC. The function $p(s|\vec{x})$ is the likelihood
\begin{equation}\label{eqn:likelihood}
p(s|\vec{x}) = C\,e^{-\left<s-h\left(\vec{x}\right)|s-h\left(\vec{x}\right)\right>/2},  
\end{equation}
where $C$ is a normalization constant.  This jump is then accepted with probability $\alpha = \min(1,H)$, otherwise the chain stays at $\vec{x}$. 

If the proposal distribution was independent of the step number, the algorithm would be Markovian. However, the MHMC algorithm we employ is not, since we use a variety of purposely directed proposal distributions that allow a range of jumps of different size and type in the parameter space. We also implement several annealing schemes to encourage movement of the chain and use time sliding to search automatically over the plunge time (which we use as a parameter instead of the initial frequency, $\nu_0$). We are using the MHMC algorithm as a search code to find the unknown parameters of the signal. Making the search non-Markovian allows more rapid convergence to the source parameters, at the expense of no longer being able to construct the posterior from the chain state distribution. In a future implementation of the pipeline, we will include a final Markov Chain stage to recover the posterior once the source parameters have been approximately identified. This was not implemented for the searches described here.

\subsection{Proposal distributions}
Our primary proposal distribution is a multi-variate Gaussian, which is constructed as a product of Gaussian distributions in each eigendirection of the Fisher Information Matrix (FIM), $\Gamma_{ij}$. The distribution in each eigendirection is taken to have zero mean and a standard deviation of $\sigma_{i} = 1/\sqrt{DE_{i}}$.  Here $D$ is the dimensionality of the search space (in this case, $D=13$, as we use normalized search templates) and $E_{i}$ is the corresponding eigenvalue of $\Gamma_{\mu\nu}$. The FIM is computed numerically, using the same truncated AK model that we employ as the search template, but with the additional simplification that the detector response is modelled using the low-frequency approximation (as in the original AK paper~\cite{BC04}), rather than via the full TDI response.

While the FIM based proposal is used at most steps, we also periodically draw a proposed point from one of several other proposal distributions:
\begin{itemize}
\item {\it Scaled uniform jumps} --- the proposal distribution is taken to be uniform within the priors for each parameter. This forces the chain to explore other, widely separated, regions of the parameter space.
\item {\it Scaled FIM jumps} --- this proposal is based on the FIM as for the standard proposal, but the proposed jumps are artificially reduced in size by a factor of ten. These proposals are almost always accepted, forcing the chain to move slightly away from secondaries and hence encouraging movement.
\item {\it Antipodal sky position} --- at low frequencies, the sky positions $(\beta, \phi_S)$ and $(- \beta, \phi_S \pm \pi)$ are equivalent in terms of the detector response. In searches for black hole binaries, it was found that the chains could often become locked on the wrong sky position, so we include a proposal that moves the chain to the antipodal sky position to avoid this problem.
\item {\it Intrinsic/extrinsic/phase only jumps} --- the waveforms can be divided into {\it intrinsic} parameters ($m$, $M$, $\nu_0$, $e_0$, $\chi$, $\lambda$), {\it extrinsic} parameters ($D_L$, $\beta$, $\phi_S$, $\theta_K$, $\phi_K$) and {\it phase offsets} ($\Phi_0$, $\gamma_0$, $\alpha_0$). Intrinsic parameters affect the frequency and phase evolution of the different harmonics, while extrinsic parameters only affect how this is projected into a detector response, and the phase offsets only define the relative phase of the different harmonics at one fiducial time. Proposing jumps in extrinsic/phase offset parameters only was designed to improve the fit while keeping the harmonic frequencies fixed (see Section~\ref{harmident}).
\end{itemize}
We note that no matter which proposal we use, all the waveform parameters are updated, i.e., we do not use Gibbs sampling to update one parameter at a time. Typically, we draw from each of the alternative proposals every few tens of chain points, but these sampling frequencies are tunable parameters which we have not yet optimised for the EMRI search, and so we do not quote specific numbers here.

\subsection{Simulated, Thermostated and Time Annealing.}
As the likelihood surface for EMRIs is full of secondary maxima, it is very easy for the search chains to get stuck. To ensure acceptable movement in the chain, we use simulated annealing to heat the likelihood surface~\cite{en2,en1}.  This smooths and flattens any bumps on the likelihood surface, which ensures the chain has a greater chance of escaping secondaries and moving uphill towards the primary peak.  In the definition of the likelihood, Eq.~(\ref{eqn:likelihood}), there is a factor of 2 in the denominator of the exponent.  We replace this factor with a ``heat'' parameter $\Theta$, defined by
\begin{equation}
\Theta = 2 \times \left\{ \begin{array}{ll} 10^{\xi\left(1-\frac{i}{T_{c}}\right)} & 0\leq i\leq T_{c} \\ \\ 1& i > T_{c}  \end{array}\right.,
\label{heatsch}
\end{equation}
where $\xi > 0$ is the heat-index defining the initial heat, $i$ is the step number of the chain and $T_{c}$ is the cooling schedule (i.e., the number of chain steps over which the cooling takes place). As the success of the choice of initial heat is only known a posteriori, we are still investigating the optimal heating scheme.

To mitigate potential problems,  we also employed a thermostated heating scheme as used in~\cite{en1} . We define
\begin{equation}
\delta = 2 \times\left\{ \begin{array}{ll} 1 & 0\leq {\rm SNR}\leq \rho_c \\ \\ \left(\frac{\rm SNR}{\rho_c}\right)^{2} & {\rm SNR} > \rho_c \end{array}\right. 
\end{equation}
and take the heat to be the maximum of $\delta$ and $\Theta$ defined by Eq.~(\ref{heatsch}). The thermostated heating scheme pumps heat into the system once we reach ${\rm SNR}=\rho_c$.  This should make the chain move uphill faster, once we have begun to match the signal.  This scheme is run in conjunction with the simulated annealing.  While we continue to calculate both heats, we always use the heat that is highest in the search. For this study we chose $\rho_c=15$.

The final annealing scheme we employ is time annealing, which is similar to the frequency annealing described in Ref.~\cite{en1}. The cost of evaluating the likelihood and the FIM for the proposal distribution depends on the length of the waveform template being used. It is inefficient to use full length (two year) templates initially, when the parameter values are poorly constrained. We therefore start off with shorter templates and increase the length of the template, $t_{\rm dur}$, as the chain progresses
\begin{equation}
t_{\rm dur}= \left\{ \begin{array}{ll} t_{\rm max} \left(t_{\rm min}/t_{\rm max}\right)^{\left(1-\frac{i}{N_{f}}\right)}& t< t_{\rm max} \\ \\ t_{\rm max} & t \geq t_{\rm max}  \end{array}\right.,
\end{equation}
where $i$ is the number of steps in the chain, $N_{f}$ is the total number of iterations in the time annealing chain, $t_{\rm min}$ is the minimum length of template and $t_{\rm max}$ is the maximum length of the template. The aim of the time annealing is to use shorter templates while the chain is taking big jumps exploring the parameter space, and then to increase the template length and refine the parameter determination once the chain has settled in the vicinity of the true parameters. For this analysis, we used $t_{\rm min} = 4$ months to ensure a reasonable SNR ($\geq 20$) in the shortest templates employed, and we took $N_f=10,000$ and $t_{\rm max}=1$yr, half the length of the MLDC data release.

In our implementation of the search, we started by running a $5,000$ iteration chain with $t_{\rm dur}$ fixed at four months, but using simulated annealing and thermostated heating. After $5,000$ iterations we began a 10,000 iteration time-annealing scheme, using thermostated heating  in conjunction but not simulated annealing. Once the time-annealing phase was complete, we cooled the surface down over 5,000 iterations during a final simulated annealing phase according to the scheme detailed by Equation~(\ref{heatsch}) without thermostated heating. We then allowed the chain to run as a standard MCMC with unit heat for 80,000 iterations to obtain a final refinement of the parameters.

\subsection{Harmonic identification}
\label{harmident}
\begin{figure}
\begin{tabular}{c}
\includegraphics[height=3.5in, width=\textwidth, angle=0]{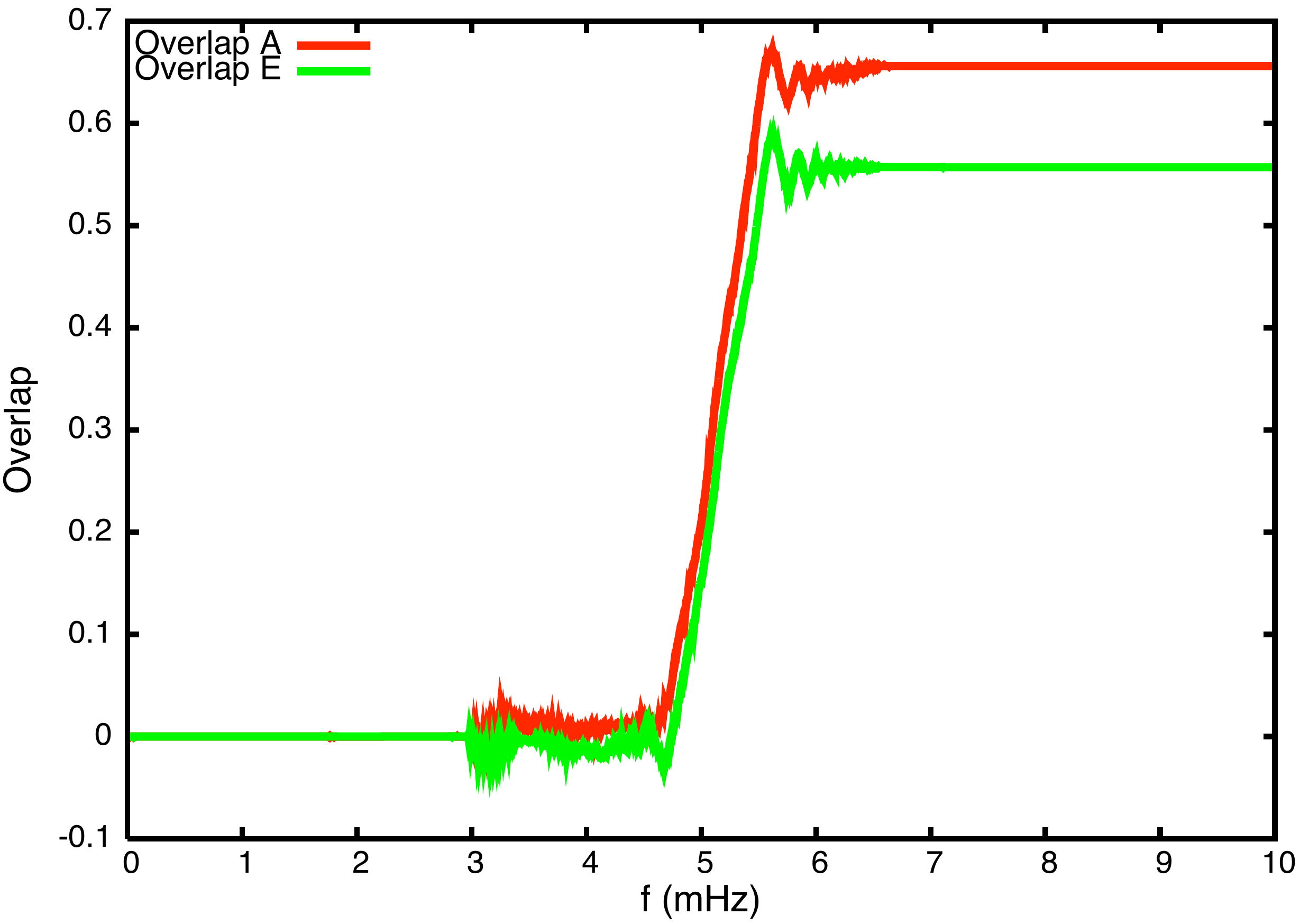}\\\includegraphics[height=3.5in, width=\textwidth, angle=0]{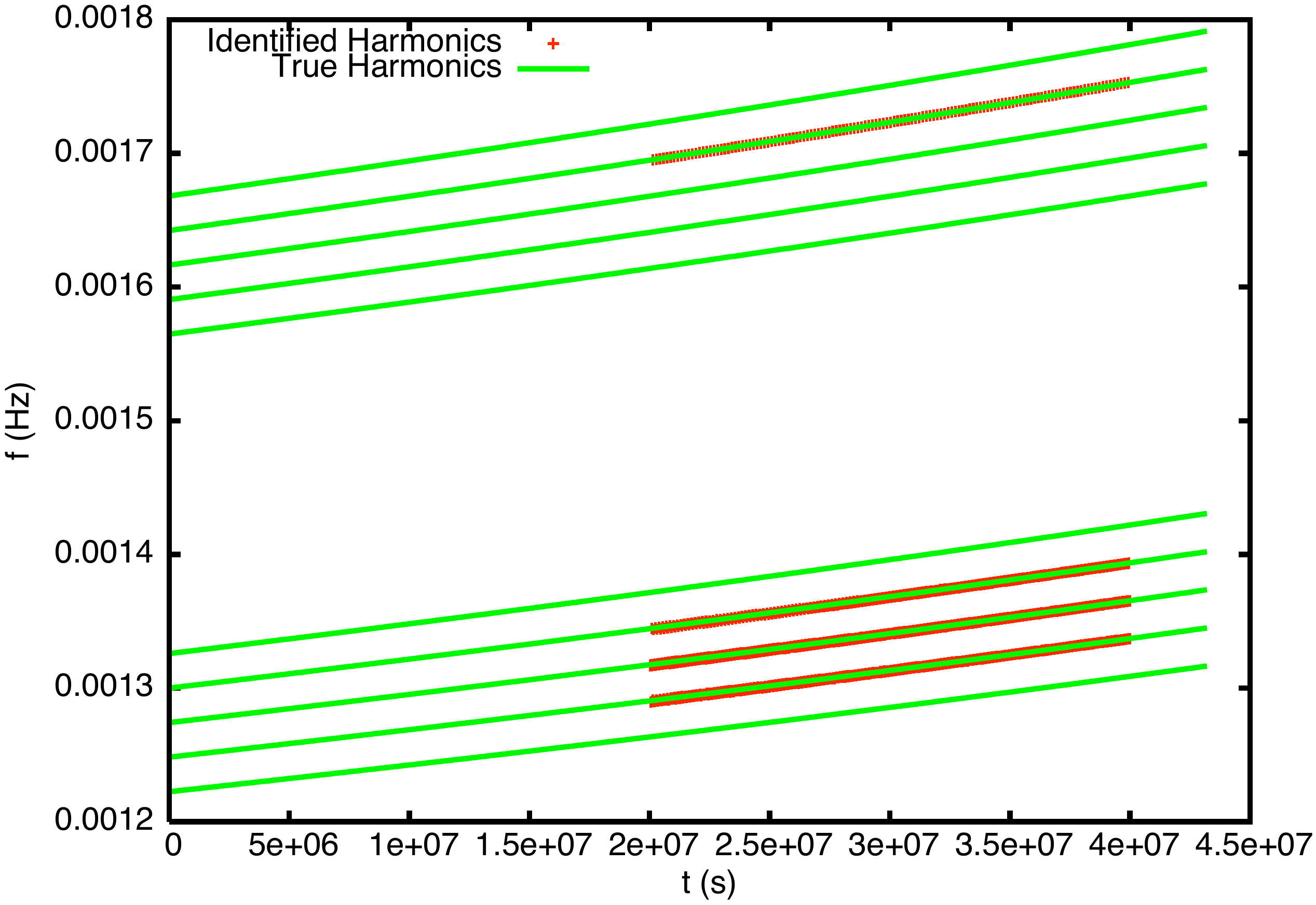}
\end{tabular}
\caption{\label{harmmatch} {\it Top panel} Cumulative overlap of a given harmonic with the data stream, as a function of frequency. The curves show the overlap of the A and E TDI data streams respectively. The overlap accumulates when the harmonic frequency is $4.5 {\rm mHz} \lesssim f \lesssim 5.5 {\rm mHz}$, which indicates the time range where it matches a harmonic of the true signal. {\it Bottom panel} True harmonics (lines) and identified harmonics (crosses) for the MLDC Round 1B training source 1B.3.2.}
\end{figure}

The EMRI likelihood space is very complicated and is characterized by many secondary maxima. It is consequently very easy for the chain to get stuck on a secondary. This was a major concern in our analysis of the MLDC Round 2 data, and again in the initial analysis of the Round 1B data. A chain started at a random point would lock very quickly onto a secondary of the signal, but would then not move far away from that point. However, chains seeded at different initial points would lock onto different secondaries. A secondary typically has one or two harmonics that overlap with harmonics of the true signal for a section of the data stream, and we can use that information to determine the true parameters. Given the parameters of a secondary, one can construct the cumulative overlap (in the frequency domain) of each harmonic of the secondary with the data stream. If the overlap is significant, the frequency range in which the overlap accumulates indicates the section of that secondary harmonic which matches a harmonic of the true signal. This is illustrated in the top panel of Fig.~\ref{harmmatch}.

In order to exploit this information, as a preliminary stage of the search we ran several ($\sim20$) chains, seeded at different points in parameter space, stopped the chains after $\sim 1000$ steps and then analysed the harmonic content of the highest SNR point found in each chain. Different chains would typically find different harmonics, and so combining this information allows a picture of the signal structure to be constructed. This is illustrated in the bottom panel of Fig.~\ref{harmmatch}. This information can be used in the search in several ways --- as a veto of secondaries (any parameter space point that does not match these harmonics cannot be the primary maximum); to fold into parameter priors (reject proposed steps that move harmonics too far away from these tracks); or as a constraint when proposing steps for the chain.

\subsection{Constrained jumps}
It is possible to modify the proposal distribution to ensure certain constraints are satisfied, e.g., the frequency of a given harmonic at a given time. Given a set of constraints, $\{f_i (\vec{x}) = 0\}$ for $i=1,\cdots,N$, we can define a set of unit vectors orthogonal to the constraint surfaces, $\vec{e_i} = \nabla f_i/|\nabla f_i|$, and decompose a step in parameter space, $\delta\vec{x}$, into a piece that maintains the constraints, $\delta \tilde{\vec{x}}$, and a piece in the space spanned by the $\vec{e_i}$,  $\delta\vec{x}=\delta \tilde{\vec{x}}+a_i\vec{e_i}$. The template mismatch, ${\cal M}$, at the new point for a jump that maintains the constraints is therefore given in terms of the FIM, $\Gamma_{ij}$, by
\begin{eqnarray}
2\,{\cal M} &=& \Gamma_{ij} \delta\tilde{x^i} \delta\tilde{x^j} = \tilde{\Gamma}_{ij}  \delta{x^i} \delta{x^j} \nonumber \\ &=& \left(\Gamma_{ij} - 2 \Gamma_{mj} C_{lk} e_k^i e_l^m + \Gamma_{mn} C_{lk} C_{pq} e_k^i e_q^j e_l^m e_p^n \right) \delta{x^i} \delta{x^j}
\label{constrFIM}
\end{eqnarray}
in which superscripts indicate vector components, $C_{ij} = (A^{-1})_{ij}$ for $A_{ij} = \vec{e_i} \cdot \vec{e_j}$ and repeated indices are summed as usual. The matrix $\tilde{\Gamma}_{ij}$ can be used in place of $\Gamma_{ij}$ to generate proposed steps that maintain the constraints. The matrix $\tilde{\Gamma}_{ij}$ has $N$ zero eigenvalues with corresponding eigenvectors $\{\vec{e_i}\}$, which are ignored when constructing the proposed step.

We implemented constrained jumps in the search, using the results of the harmonic analysis described in Section~\ref{harmident} to determine the frequencies of several harmonics and their rates of change at a specified time. In practice, we found it best to use the frequencies of three harmonics, and the time derivative of the dominant harmonic as the constraints. Additional frequencies or derivatives did not provide any extra information. For the AK model, it is easy to translate such constraints into parameter values and hence reduce the six-dimensional intrinsic parameter space --- $M, m, S, \lambda, e_0, \nu_0$ --- to a two-dimensional space spanned by $e_0$ and $\lambda$. We actually used this dimensional reduction in our implementation of the search, although the general expression~(\ref{constrFIM}) will be necessary for a generic waveform model.

\begin{figure}
\begin{center}
\includegraphics[keepaspectratio=true, width=5.2in, angle=0]{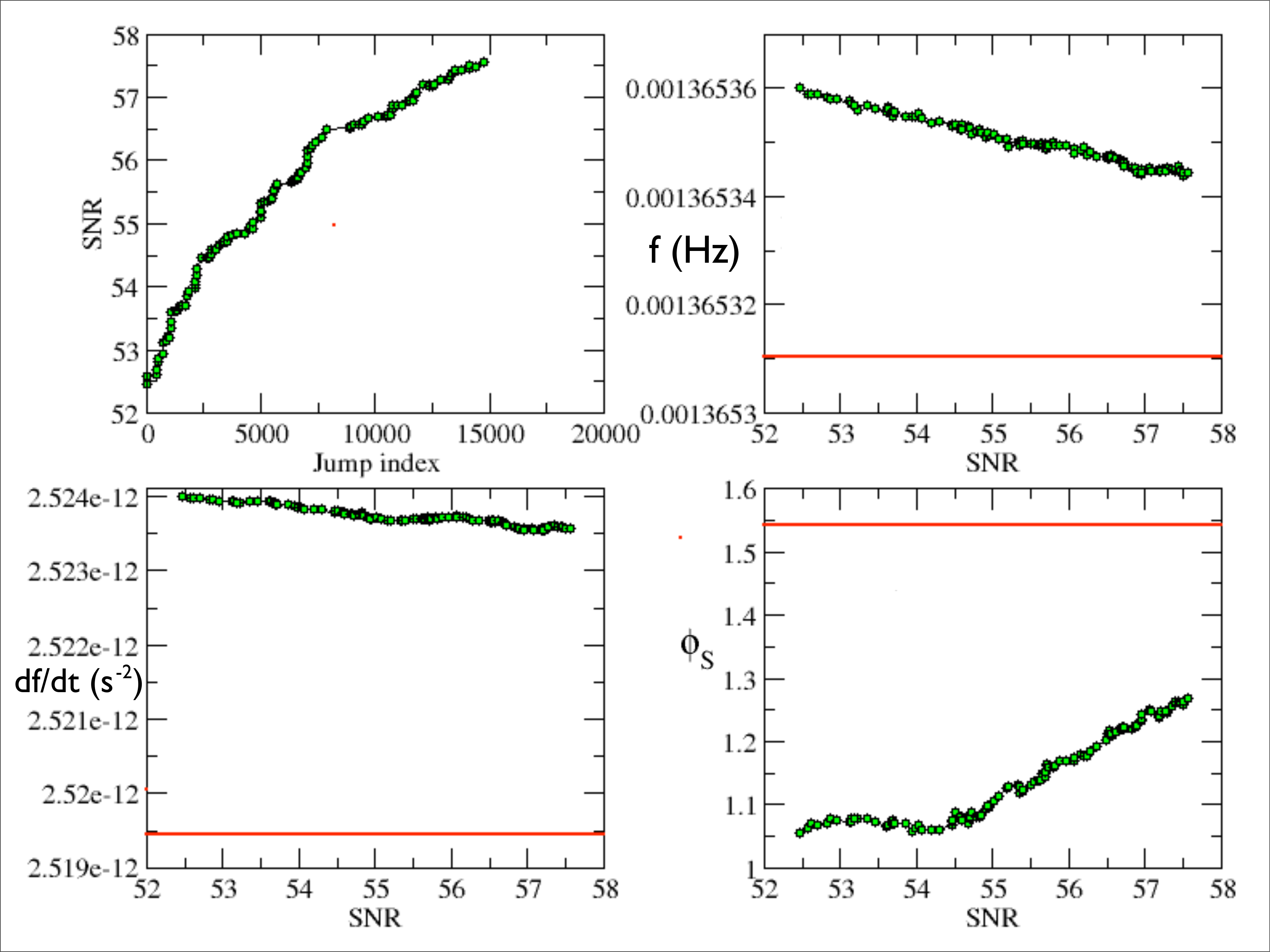}
\end{center}
\caption{\label{semiconsrch} SNR as a function of chain index (top-left); frequency of dominant harmonic  vs. SNR (top-right); rate of change of frequency of dominant harmonic vs. SNR (bottom-left); and sky longitude vs. SNR (bottom-right), in a partially constrained search for the MLDC Round 1B.3.2 challenge source. The chain is moving steadily towards the true parameter values, indicated by horizontal lines. This run did not employ either simulated annealing or time-annealing.}
\end{figure}

The constrained search was able to rapidly improve the SNR from the initial guess and if the constraints were specified exactly (using the MLDC training data), the chain moved steadily to the correct point. However, the harmonic analysis was not able in general to determine precise values for the frequencies. An error of $\sim 10^{-7}\, {\rm Hz}$ in one frequency leads to a dephasing after $\sim 4$ months, which limits the usefulness of the fully constrained search. We found it more effective to use a partially constrained search, i.e., using the harmonic analysis to estimate frequencies and frequency derivatives at a certain time {\it with estimated uncertainties}. It is possible to reparameterise the waveform in terms of the values of the three fundamental frequencies at a certain time, plus the time derivative of the dominant harmonic at the same time. An MHMC chain can then be run on this alternative parameter space, using tight priors on those frequencies assigned using the frequency uncertainty estimates. This can also be done directly in the physical parameter space by using the constrained FIM, $\tilde{\Gamma}_{ij}$, and additionally allowing small jumps in the directions $\vec{e_i}$ orthogonal to the constraint surface. The results of one partially constrained search are illustrated in Figure~\ref{semiconsrch}. This type of search was found to be much more effective, although our initial implementation of the proposal was inefficient which led to a low acceptance rate.

\section{Round 1B Results}\label{results}
At the time of the MLDC submission deadline (December 2007), we were still developing the details of the search outlined above, and therefore just submitted the maximum a posteriori (MAP) parameters that had been found in the search by that time. We used the MAP values as we were sure at this stage that we were either still moving towards the true solution, or stuck on a secondary.  It therefore made no sense to use the mean of a posterior to determine the best fit paramteres.  Our strategy involved starting 10 chains at different random start points for each challenge.  Due to time constraints we needed to stop the MCMC chains after between 35,000 and 40,000 iterations out of 80,000.   We used the MAP values from each of these chains as starting points for the harmonic identification and constrained stages of the search.  The initial MHMC runs took 2-3 weeks each, the harmonic identification was then quick (a few hours), but the final constrained stage also took several weeks. We expect these run times to decrease as the algorithm is refined, but a final run time of $\sim1$ week is a reasonable guess. We should mention that the search of Challenge 1B.3.2 was the most advanced at the time of submission, and this was the only source for which we had begun the constrained jump phase of the search by that time.

A harmonic analysis indicated that we were in the vicinity of secondary maxima for sources 1B.3.1, 1B.3.3, 1B.3.4 and 1B.3.5. The harmonics of our best guess parameters for 1B.3.2 appeared to be approximately in the right place, although the total SNR of this point did not lie within the MLDC prior, indicating that the parameters were not completely correct. The submitted results are summarised in Tables~\ref{restab}--\ref{restabB}, along with the true values of the Challenge parameters.

As expected, our results for 1B.3.2 agreed quite well with the true parameters, but in the other cases we were quite far off. The recovered parameters correspond to secondaries of the likelihood in all cases, which share harmonics with the true signal. For all sources except 1B.3.1, we appear to have locked onto the antipodal sky location, despite the inclusion of a sky position flip as one of the proposal distributions. We hope to avoid this in the future by implementing {\it delayed rejection}~\cite{neilemri} which will be discussed in Section~\ref{fut}.

\begin{table}
\begin{tabular}{|cc|c|c|c|c|c|c|}
\hline
\multicolumn{2}{|c|}{Source}  & $S/M^2$&$m/M_{\odot}$&$M/(10^6\,M_{\odot})$&$\nu
_0$(mHz)&$e_0$&$\lambda$\\
\hline
{\bf 1B.3.1}&{\em True}&0.69816& 10.296& 9.5180& 0.19204& 0.21438& 0.43946\\ 
&{\em Recovered}& 0.63624&10.500&10.359&0.18711& 0.15810&0.50800\\ \hline
{\bf 1B.3.2}&{\em True}&0.63796&9.7711& 5.2156& 0.34228&0.20791&1.4358\\ 
&{\em Recovered}& 0.63971&9.7751&5.2076&0.34223&0.20941&1.4399\\ \hline
{\bf 1B.3.3}&{\em True}&0.53326&9.6973&5.2197& 0.34257& 0.19927&0.92822\\ 
&{\em Recovered}& 0.59655&10.193&5.2344&0.34236&0.19647&0.75882\\ \hline
{\bf 1B.3.4}&{\em True}&0.62514&10.105&0.95580& 0.85144& 0.45058&1.6707\\
&{\em Recovered}& 0.63104&10.085&1.0439& 0.79510&0.44077&2.1837\\ \hline 
{\bf 1B.3.5}&{\em True}&0.65830&9.7905&1.0334& 0.83218& 0.42691&2.3196\\
&{\em Recovered}&0.67701&9.8849&0.97872&0.83390&0.42950&2.5092\\
\hline
\end{tabular}
\caption{Comparison of best recovered parameters at time of MLDC submission to true parameters.}
\label{restab}
\end{table}

\begin{table}
\begin{tabular}{|cc|c|c|c|c|}
\hline
\multicolumn{2}{|c|}{Source}  & $\beta$&$\phi_S$& $\theta_K$&$\phi_K$ \\
\hline
{\bf 1B.3.1}&{\em True}&0.55258 &4.9104& 1.7625& 2.0472\\ 
&{\em Recovered}&0.64558&4.9473&2.2005&1.7141\\ \hline
{\bf 1B.3.2}&{\em True}&0.35970&4.6826&3.0372&4.0382\\ 
&{\em Recovered}&-0.87434&1.1145&1.5836&6.0901\\ \hline
{\bf 1B.3.3}&{\em True}&0.98166&0.70967&1.5364&4.1487\\ 
&{\em Recovered}&-0.69482&3.8708&2.2751&3.8650\\ \hline
{\bf 1B.3.4}&{\em True}&-0.98020&0.97873&1.7601&4.1164\\
&{\em Recovered}&0.81338&3.2806&0.60980&6.0855\\ \hline 
{\bf 1B.3.5}&{\em True}&-1.1092&1.0876&0.84039&5.7564\\
&{\em Recovered}&0.39053&1.9690&1.9561&5.1293\\
\hline
\end{tabular}
\caption{Comparison of best recovered parameters at time of MLDC submission to true parameters --- sky position and source orientation.}
\label{restabB}
\end{table}

\section{Summary and future plans}\label{fut}
Our final search algorithm for identifying isolated EMRIs in instrumental noise was as follows: 1) run a set of chains without annealing for a small number of chain steps; 2) identify harmonics of the true signal from the highest SNR points found in the preliminary chains; 3) run a partially constrained search, including temperature and time annealing, to refine the source parameters. The search runs were incomplete at the time of the MLDC deadline, but subsequent runs have moved closer to what we now know are the true parameters. The convergence rate is slow, however, so we are currently exploring several improvements to the algorithm, including
\begin{itemize}
\item {\it Fast waveform model.} By using the low-frequency approximation to the detector response, and interpolation of the barycentre waveform, it is possible to speed up the waveform and FIM evaluations which are the bottlenecks in the current code.
\item {\it Semi-coherent analysis.} As considered for template based searches in~\cite{gairetal}, by dividing the data stream into sections of a few months in length, and searching these separately in parallel, the search speed is increased. The main difficulty is forcing parameter consistency between the different segments.
\item {\it Delayed rejection.} This is a technique whereby the chain is forced to accept a (large) jump in the parameter space, but the likelihood at the original point, $\vec{x_i}$, is recorded. The chain is then allowed to run (with small jumps) for a pre-specified number of steps, e.g., 100, to reach a final point, $\vec{x_f}$, before the Metropolis-Hastings ratio is evaluated using $\vec{x_i}$ and $\vec{x_f}$. If this step is rejected, the chain goes back to $\vec{x_i}$ and another jump is proposed. This technique can help the chain jump to widely separated secondaries in the parameter space, and it might also encourage the chain to switch to the correct antipodal sky position, as mentioned earlier.
\item {\it Numerical F-statistic} It is relatively straightforward to construct a proposal that changes the intrinsic parameters of the source in order to achieve a rotation of the signal harmonics, i.e., to keep the dominant harmonic in the same place, but change the indices $(n,l,k)$ of that harmonic. However, it will usually be necessary to change the extrinsic parameters and phase angles in order to find a high SNR at the new point.  This can be accomplished by proposing a jump in the intrinsic parameters, and then finding the best-fit extrinsic parameters at the new point, either by using a template grid or by using a mini ($\sim100$ iteration) monte carlo chain that explores only the extrinsic parameters. The likelihood maximized over extrinsic parameters can then be used to evaluate the Metropolis-Hastings ratio for the proposed jump in intrinsic parameters.
\end{itemize}

A short-coming of the algorithm described here will become apparent when the data stream becomes more complex and contains multiple sources, as in MLDC Round 3. When multiple sources are present, distinguishing between a weak harmonic that is a side-band of an identified bright source and one that is the dominant harmonic of a second, weaker signal, is difficult. The location in time and frequency, and the shape of the track, will be useful diagnostics, but some confusion will be inevitable. As an alternative to identifying harmonics to use as constraints, an understanding of how harmonic rotations/shifts can be achieved by parameter changes should allow the formulation of proposal distributions that will move the chain from one secondary to another (similar to the ``island hopping'' used in supermassive black hole binary searches~\cite{en1}). The aim of such a proposal would be to make the chain jump between points in parameter space that had harmonics in common with the true signal and hence the chain should find the true parameters more quickly. Such proposals will be explored when we apply these algorithms to the Round 3 data.

\ack JG acknowledges support from the Royal Society and thanks the Albert Einstein Institute for hospitality and support while this work was being completed. The work of SB and EKP was supported by DLR (Deutsches Zentrum f\"ur Luft-  und Raumfahrt). LB acknowledges support from PPARC/STFC through grant number PP/D00111011.

\pagebreak
\appendix


\section*{References}
\begin{thebibliography}{99}
\bibitem{gairetal} Gair~J~R, Barack~L, Creighton~T, Cutler~C, Larson~S~L, Phinney~E~S \&
Vallisneri~M, Class. Quant. Grav. {\bf 21} S1595 (2004).

\bibitem{wengair05} Wen~L \& Gair~J~R, Class. Quantum Grav.~{\bf 22} S445 (2005).

\bibitem{gairwen05} Gair~J~R \& Wen~L, Class. Quantum Grav.~{\bf 22} S1359 (2005).

\bibitem{wcg06} Wen~L, Chen~Y \& Gair~J~R, AIP Conf. Proc.~{\bf 873} 595 (2006).

\bibitem{gairjones06} Gair~J~R \& Jones~G, Class. Quantum Grav.~{\bf 24}  1145 (2007).

\bibitem{proc13} Gair~J~R, Mandel~I \& Wen~L, Journal of Physics Conf. Proc., accepted. Preprint arXiv:0710.5250 (2007).

\bibitem{andrieu} Andrieu~C and Doucet~A, IEEE Trans. Signal Process. {\bf 47} 2667 (1999).

\bibitem{umstat} Umstatter~R, Christensen~N, Hendry~M, Meyer~R, Simha~V, Veitch~J, Vigeland~S \& Woan~G, Phys. Rev. D {\bf 72} 022001 (2005).

\bibitem{cc05} Cornish~N~J \& Crowder~J, Phys. Rev. D {\bf 72} 043005 (2005).

\bibitem{cc2} Crowder~J \& Cornish~N~J, Phys.~Rev.~D {\bf 75} 043008 (2007).

\bibitem{en3} Cornish~N~J \& Porter~E~K, Class. Quant. Grav. {\bf 23} S761 (2006).

\bibitem{vecchio1} Wickham~E~D~L, Stroeer~A \& Vecchio~A, Class. Quant. Grav. {\bf 23} S819 (2006).

\bibitem{en2} Cornish~N~J \& Porter~E~K, Phys. Rev. D {\bf 75} 021301 (2007).

\bibitem{en4} Cornish~N~J \& Porter~E~K, Class. Quant. Grav. {\bf 24} S501 (2007).

\bibitem{en1} Cornish~N~J \& Porter~E~K, Class.~Quant. Grav.~{\bf 24} 5729 (2007).

\bibitem{stroeer06} Stroeer~A, Gair~J~R \& Vecchio~A, AIP Conf. Proc., {\bf 873} 444 (2006).

\bibitem{mldc1} Arnaud~K~A et al., AIP Conf. Proc., {\bf 873} 619 (2006).

\bibitem{neilemri} Cornish~N~J, Preprint arXiv:0804.3323 (2008)

\bibitem{mldc2} Babak~S, Baker~J~G, Benacquista~M~J, Cornish~N~J, Crowder~J, Cutler~C, Larson~S~L, Littenberg~T~B, Porter~E~K, Vallisneri~M \& Vecchio~A, {\emph Proceedings of the 7th Amaldi Conference on GWs, July 2007, Sydney, Australia}. arXiv:0711.2667 (2007)

\bibitem{mldc3} The MLDC taskforce, {\emph Proceedings of the 12th GWDAW, December 2007, Boston, USA}. 

\bibitem{BC04} Barack~L \& Cutler~C, Phys. Rev. D~{\bf 69} 082005 (2004).
\end{thebibliography}
\end{document}